\documentstyle[twoside, epsf]{article}

\input ibvs2.sty

\begin{document}
\IBVShead{6089}{4 January 2014}

\IBVStitle{V1117 H{\lowercase{er}}: A Herbig A{\lowercase{e}} star at high 
Galactic latitude?}

\IBVSauth{Kun, M.; R\'acz, M.; Szabados, L.}

\IBVSinsto{Konkoly Observatory, H-1121 Budapest, Konkoly Thege \'ut 15--17, Hungary, e-mail: kun@konkoly.hu}

\SIMBADobj{V1117 Her}
\IBVSabs{We examine the long-term light curve, optical spectrum, spectral} 
\IBVSabs{energy distribution, and Galactic location of V1117 Her in order}
\IBVSabs{to establish its nature. V1117 Her is most probably a young }
\IBVSabs{intermediate-mass star whose cyclic brightness dimmings are caused} 
\IBVSabs{by changing circumstellar dust structures.}

\begintext

\section{Introduction}

The variations of the star located at 
RA(2000) = 16$^{\mathrm{h}}39^{\mathrm{m}}06.42^{\mathrm{s}}$, D(2000)=$+09^{\mathrm{o}}47^{\prime}55.3^{\prime\prime}$ 
were discovered by Blazhko (1929), but the nature of the variability has 
remained unexplored until recently. Thanks to the massive photometric 
monitoring projects {\it ROTSE\/} (Akerlof et~al. 2000) and {\it ASAS\/} 
(Pojmanski 2002), and the observers of the {\it AAVSO\/} now decade-long 
{\it V\/}-band light curve is available for this star. 
The Northern Sky Variability Survey ({\it NSVS\/}, Wo\'zniak et~al. 2004a), conducted in the course of the Robotic Optical Transient Search Experiment (ROTSE-I), catalogued NSVS\,1639065+094755 as a long period variable having an (unfiltered) average magnitude of 13.080, amplitude of 1.938 mag, and period of 114~days (Wo\'zniak et~al. 2004b). The {\it support vector machines\/} method, applied during the construction of the {\it NSVS\/} catalogue, classified the star as a Mira variable, combining its light curve characteristics with colour indices formed from the median {\it ROTSE\/} and {\it 2MASS\/} magnitudes. However, no spectroscopic observation, required for the confirmation of this classification is available in the literature.
The name V1117~Her was given in 2008 (Kazarovets et~al. 2008). The GCVS 
(Samus et~al. 2007--2012) classification of this star is {\it IS\/}, i.e. 
``rapid irregular variables having no apparent connection with diffuse 
nebulae and showing light changes of about 0.5--1.0~mag within several hours 
or days''. 
The long-term light curve of V1117~Her exhibits steep fadings resembling those of the young UX~Ori-type stars. This may be a reason that this star has been included in the young stellar object (YSO) monitoring programme of the {\it AAVSO\/} since 2006. The data are available upon request at the {\it AAVSO\/} Web site.\footnote{\tt http://www.aavso.org/} UX~Ori-type variables belong to the class of Herbig~Ae stars, i.e. intermediate-mass pre-main sequence stars surrounded by circumstellar dust disks and/or envelopes. Their recurrent dimmings most probably originate from the 
variations of circumstellar extinction due to non-axisymmetric and changeable, 
orbiting dust structures (e.g. Bibo \& Th\'e 1990, Herbst et~al. 1994). 
The high infrared brightness of V1117~Her at the {\it 2MASS\/} and {\it WISE\/} wavelengths indeed suggests the presence of circumstellar dust. This star, however, lies at a high Galactic latitude ($b$=+33.80), far from known star forming regions. The interstellar extinction toward its direction, read from the SFD (Schlegel et~al. 1998) Galactic dust map, is $A_{\mathrm V} \sim 0.07$~mag. To test the hypothesis that V1117~Her is a pre-main sequence star, in this paper we examine its brightness and colour variations, optical spectrum, spectral energy distribution and the possible birthplace. 

\section{Data}

\subsection{Light curves}

The {\it V\/}-band light curve, containing the 179 {\it NSVS\/} and 151 {\it ASAS\/} 
data points, as well as 303 {\it AAVSO\/} CCD  and 271 {\it AAVSO\/} visual 
measurements, is shown in Fig.~1. In all, the 633 {\it V}-band CCD and 271 
visual measurements cover a time span of 5231 days (14.33 years). Since the {\it NSVS\/} photometry is tied to the Johnson {\it V\/} magnitude scale (Wo\'zniak et~al. 2004a), we plotted the {\it NSVS\/} magnitudes together with the {\it V\/}-band data.
The average of the 633 CCD measurements is $\langle V \rangle = 13.223$~mag. 
The peak-to-peak {\it V}-band amplitude is 2.775~mag. 

\IBVSfig{7cm}{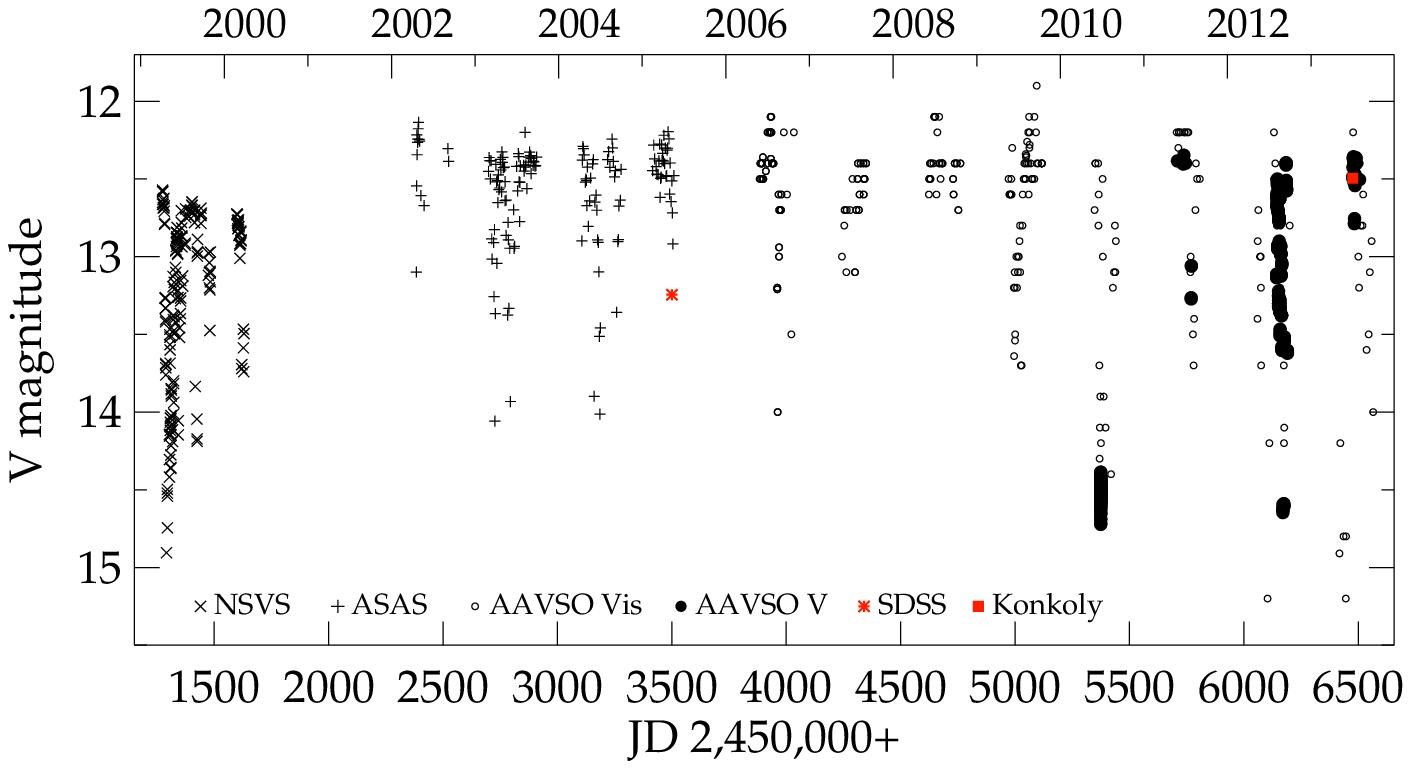}{V-band light curve of V1117 Her, based on all of the available data.}
\IBVSfigKey{6089-f1.ps}{V1117 Her}{light curve}

The light curve in Figure~1 clearly indicates the presence of cyclic fading events. The visual inspection of the temporal brightness variations implies a cycle length of about 400 days. To determine a more accurate value of the cycle length, a period search was performed, using the program package MUFRAN (Koll\'ath 1990). This software calculates the Fourier transform of the photometric time series. The visual observations have not been taken into account during the period analysis. The {\it ASAS\/}, {\it NSVS\/}, and {\it AAVSO\/} {\it V\/}-band CCD data were merged into one data file containing also our single-epoch photometric {\it V\/} observation, as well as the single-epoch {\it SDSS\/} magnitude transformed to the {\it V\/} band (see Sect.~2.2). 

Because the light curve itself is neither strictly repetitive, nor sinusoidal, no high peak is expected in the noisy periodogram. Nevertheless, a clear periodicity appears at 408.247 days in accordance with the value expected from the visual inspection of the light curve in Fig.~1. 

The selected frequency peak in the periodogram (which is not the highest one) is physically meaningful because it appears in the power spectra of both {\it AAVSO\/} and {\it ASAS\/} photometric observational series when studied separately. The uncertainty of the period is estimated to be $\pm$11.5 days from the width of the selected peak in the periodogram.

The {\it V\/} data folded with the best period value of 408.247 days are plotted in Fig.~2 (left panel). The symbols are the same as in Fig.~1, and the data based on visual observations (disregarded during the period search) are also plotted here. The phase curve implies an eclipsing-like behaviour.  This phase curve is similar to that of UX~Ori, the archetype of UXor variables (Fig.~4 in Rostopchina et~al. 1999), implying eclipses. The eclipsing-like phase curve in Fig.~2 shows a secondary minimum but this feature is outlined by a few visual data only thus it may not be a real phenomenon. If it exists, it may indicate an asymmetric distribution of the circumstellar matter. The histograms of the {\it V\/} magnitudes for each data series are plotted in the right panel of Fig.~2.

Both the {\it NSVS\/} and {\it ASAS\/} histograms suggest that  the light curve is shaped by short dips, superimposed on a higher flux level. The mean brightness apparently increased between the {\it NSVS\/} and {\it ASAS\/} surveys, although 
it cannot be excluded that the differences originate from some inconsistency between the {\it NSVS\/} and Johnson {\it V\/} magnitude scale. The histogram of the {\it AAVSO\/} data is biased by the great density of measurements near the deep minimum in 2010. Both the amplitude and data distribution support the hypothesis that V1117~Her is a UX~Ori type star (cf. Herbst et~al. 1994, Xiao et~al. 2010). 

\IBVSfig{7cm}{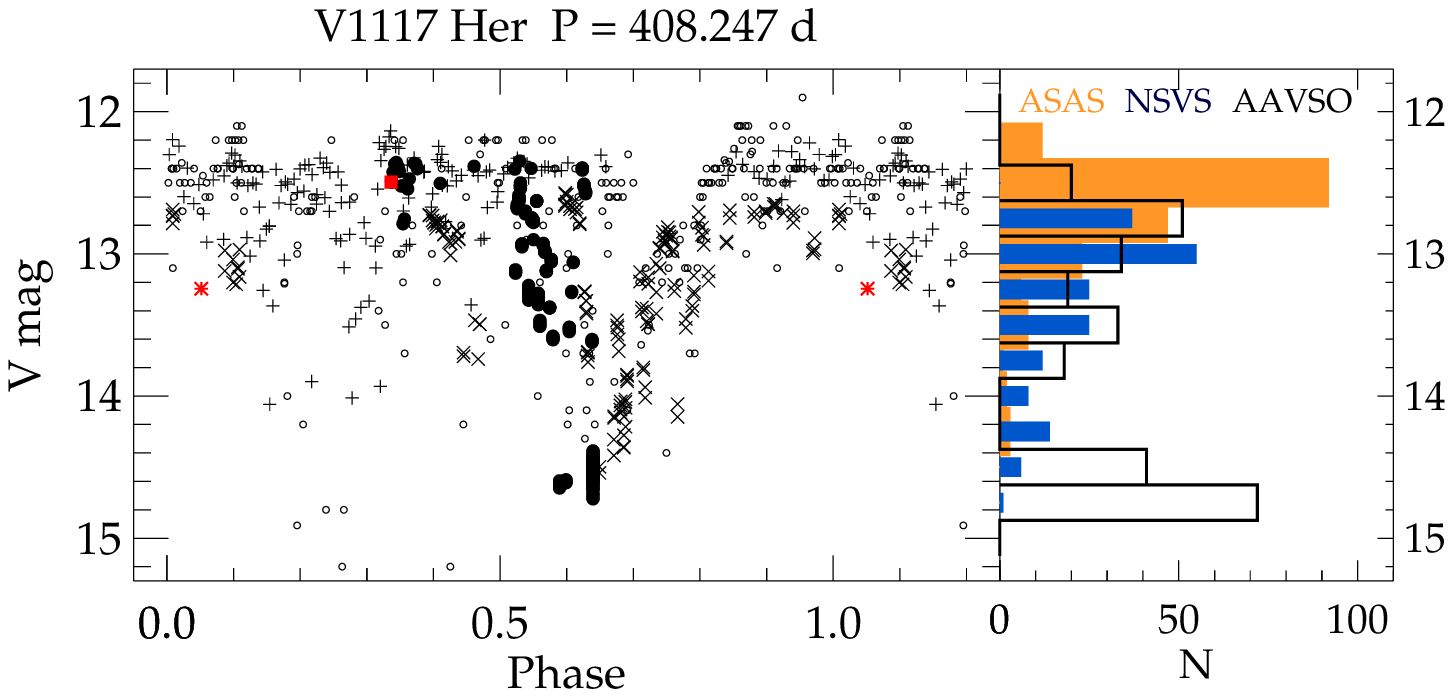}{Left: The{\it V} data folded on the best period value of 408.247 days. Symbols are same as in Fig.~1. Right: Histograms of the {\it V} magnitudes in the {\it NSVS}, {\it ASAS}, and {\it AAVSO} data bases.}
\IBVSfigKey{6089-f2.ps}{V1117 Her}{light curve}

\subsection{Colour behaviour}

The first multiband optical data of V1117~Her can be found in the {\it SDSS\/} data base (Ahn et~al. 2012), and were obtained on 2005 May 10. We transformed the {\it griz\/} magnitudes into the {\it BVR$_{\mathrm{C}}$I$_{\mathrm{C}}$} system using the formulae published by Ive\'zi\v{c} et~al. (2007).  The {\it V\/} data point obtained in this manner is also plotted on the light curve (red asterisk). 

We observed V1117~Her on 2013 July 1 in the {\it BVR$_{\mathrm{C}}$I$_{\mathrm{C}}$} bands using the 1-m RCC telescope of the Konkoly Observatory, equipped with a Princeton Instruments VersArray:1300B camera. After bias subtraction and  flat-field correction we performed aperture photometry on each star found in the images. Nineteen of these star had high-quality {\it SDSS gri} data. We calculated their  {\it BVR$_{\mathrm{C}}$I$_{\mathrm{C}}$} magnitudes as above, and used these data to establish the transformation between the instrumental and the standard {\it BVR$_{\mathrm{C}}$I$_{\mathrm{C}}$} system. The resulting magnitudes of V1117~Her are listed in Table~1, and the {\it V\/} magnitude is plotted in Fig.~1 (red filled square).

\begin{table}
{\small
\caption{{\it BVR$_{\mathrm{C}}$I$_{\mathrm{C}}$} magnitudes of V1117~Her on July 1 2013}
\begin{center}
\begin{tabular}{lcccc}
\noalign{\smallskip}
\hline
\noalign{\smallskip}
JD    &  {\it B\,($\Delta$B)} & {\it V\,($\Delta$V)}  & {\it R$_{\mathrm C}$\,($\Delta$R$_{\mathrm C}$)}  &  {\it I$_{\mathrm C}$\,($\Delta$I$_{\mathrm C}$)}  \\
\hline
\noalign{\smallskip}
2456475.4083  &   12.792\,(0.054)   &   12.493\,(0.015)   &   12.260\,(0.050)  &   12.032\,(0.051) \\
\noalign{\smallskip}
\hline
\end{tabular}
\end{center}}
\end{table}

The {\it AAVSO\/} data base contains three {\it B\/}-band measurements. Together with the {\it SDSS\/} and our own data we plotted a {\it V\/} vs. {\it B$-$V\/} colour--magnitude diagram in Fig.~3 (left panel). The slope of the interstellar extinction, assuming {\it R}$_{\mathrm{V}}=3.1$ is also indicated. The few available data points suggest that the star is reddening when fading in accordance with the interstellar extinction law, suggesting that dust structures, moving into the line of sight, may cause the light variations.

The middle panel of Fig.~3 shows the position of V1117~Her in the {\it 2MASS\/}
colour-colour diagram. We plotted for comparison three well-known UXor
variables. All of them are found to the right of the reddened main sequence,
around the  locus of unreddened pre-main sequence stars surrounded by
circumstellar disks (Meyer, Calvet \& Hillenbrand 1997).

\IBVSfig{6cm}{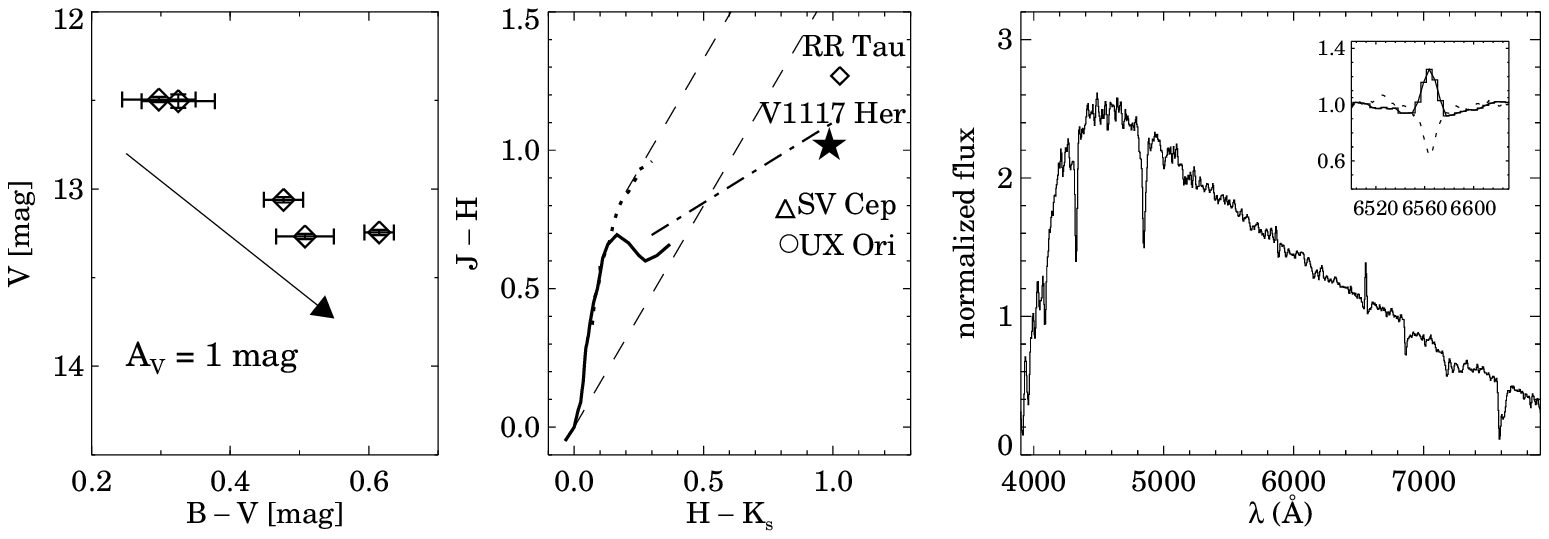}{Left: {\it V} vs. {\it B}$-${\it V} colour--magnitude diagram of V1117~Her. The arrow indicates the displacement resulting from an interstellar extinction of {\it A}$_{\mathrm{V}}=1$~mag.
Middle: Position of V1117~Her and three well-known UX~Ori type stars in the {\it 2MASS} {\it J}$-${\it H} vs. {\it H}$-${\it K}$_s$ colour-colour diagram. Solid line indicates the colours of the main sequence stars, and dotted line the giants {(Bessell \& Brett 1988)}. Dashed lines border the band of the reddened main sequence and giant stars, and the dash-dotted line is the locus of T~Tauri stars {(Meyer et~al. 1997)}. Right: Optical spectrum of V1117 Her, obtained with the low-resolution spectrograph installed on the 1-m RCC telescope of the Konkoly Observatory on 2013 July 1. The inset shows the H$\alpha$ line, together with that of the A9 type star HD~23733 (dotted line).}
\IBVSfigKey{6089-f3.ps}{V1117 Her}{other}

\section{Spectral type of V1117 Her} 

An optical spectrum of V1117~Her, centred on 5500\,\AA\ and covering a 4500~\AA-wide region was obtained on 2013 July 1 with the low-resolution spectrograph installed on the 1-m RCC telescope of the Konkoly Observatory.  The exposure time was 1800\,s. Using the 300~lines/mm grating and 3$^{\prime\prime}$ slit width, the spectral resolution was {\it R\/} = 7.3\,\AA. Spectrum of a halogen incandescent lamp was detected for flatfield correction, and a Hg--Ne lamp was observed for wavelength calibration. Neither telluric correction, nor flux calibration was applied. The spectrum was reduced in IRAF, and is shown in the right panel of Fig.~3. The strong Balmer absorption lines and weak G-band indicate a late A-type star, and the H$\alpha$ emission suggests that V1117~Her may be a pre-main sequence star. We measured the equivalent widths of the H$\beta$ and H$\gamma$ lines using the IRAF {\em sbands} task, and compared them with those in a number of standard stars both observed with the same instrument setup and found in the spectrum library of Jacoby, Hunter, \& Christian (1984). This procedure resulted in a spectral type of A8--A9. The equivalent width of the H$\alpha$ emission line is EW(H$\alpha$)=$-5.25 \pm 0.5$\,\AA. This emission is superimposed on the photospheric absorption. We measured EW(H$\alpha$)=$4.87 \pm 0.5$\,\AA\ in the A9-type star HD\,23733, thus the total EW of the emission in V1117~Her is about $-10.1\pm0.7$\,\AA. Chromospheric emission in a late A-type star is negligible, therefore the H$\alpha$ emission probably originates from accretion and wind in V1117~Her. 

\section{Spectral energy distribution}

\IBVSfig{7cm}{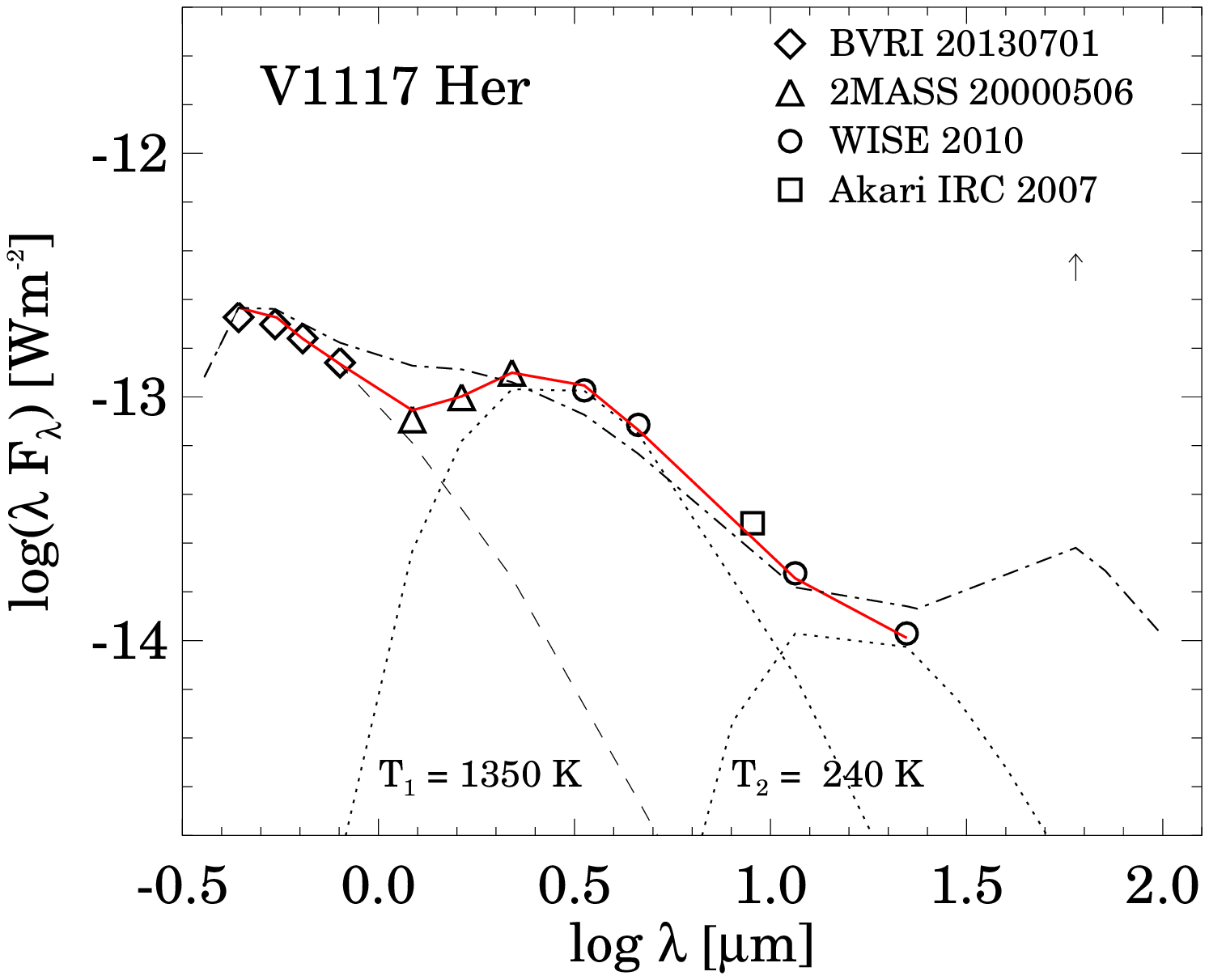}{Spectral energy distribution of V1117 Her, based on our {\it BVR\/$_{\mathrm C}$I\/$_{\mathrm C}$} data, {\it 2MASS\/} {\it JHK$_{s}$}, {\it Akari IRC\/} 9.0\,$\mu$m, and {\it WISE\/} 3.4, 4.6, 12, and 22\,$\mu$m data. Dashed line shows the photosphere of an A9 type star behind a foreground extinction of $A_{\mathrm V}=0.07$\,mag. Dotted lines show the blackbody curves fitted to the IR excess fluxes, and the solid line indicates the sum of the photosphere and the two blackbodies. The dash-dotted line indicates the YSO model No. 3015085 from the model grid of Robitaille et~al. (2007), viewed from a distance of 300\,pc, at an inclination of $87.13^{\mathrm{o}}$, and through a foreground interstellar extinction $A_{\mathrm{V}} = 0.07$\,mag.  Upward arrow indicates the {\it IRAS} lower flux limit at 60\,$\mu$m.}
\IBVSfigKey{6089-f4.ps}{V1117}{other}

The spectral energy distribution (SED) of V1117~Her, based on our {\it BVR\/$_{\mathrm C}$I\/$_{\mathrm C}$} data, {\it 2MASS\/} {\it JHK$_{s}$} (Cutri et~al. 2003), {\it Akari IRC\/} 9.0\,$\mu$m (Ishihara et~al. 2010), and {\it WISE\/} 3.4, 4.6, 12, and 22\,$\mu$m (Wright et~al. 2010) data is plotted in  Fig.~4. The dashed line indicates the photospheric flux of an A9 type star, fitted to our {\it V\/}-band data, using the colour indices of Pecaut \& Mamajek (2013), and assuming a foreground extinction $A_{\mathrm V} = 0.07$\,mag. The {\it near-infrared bump}, observed in several Herbig~Ae stars (e.g. Natta et~al. 2001), is conspicuous in the SED. It can be fitted with a 1350~K blackbody (dotted curve in Fig.~4). This temperature is close to the sublimation temperature of $\sim 1500$~K of silicate grains, suggesting that a large amount of dust is found near the sublimation radius, located at 0.3--0.5~AU from the star (Dullemond \& Monnier 2010). The 9--22\,$\mu$m fluxes suggest the presence of colder dust, located farther from the star.  

\section{Discussion}

The absolute visual magnitude of an A9 type star ($T_{\mathrm eff} = 7390$\,K, Kenyon \& Hartmann 1995) on the zero-age main sequence (ZAMS) is $M_{\mathrm V} \approx +2.4$ (e.g. Siess et~al. 2000). The mass of such a star is about 2\,M$_{\odot}$. The optical colour indices in Table~1 suggest negligible reddening. With the assumptions that V1117~Her is on the ZAMS and its light suffers an extinction of $A_{\mathrm V}=0.07$\,mag at the maximum brightness of {\it V\/}=12.135, we find that its distance from us is 860\,pc. It corresponds to a distance of $z \approx 430$\,pc from the Galactic plane, well outside the disk of the interstellar medium. If V1117~Her is above the ZAMS, even larger distances are obtained. A young star so far from the Galactic disk is fairly unlikely. This Galactic position suggests that V1117~Her is either not a young star and its Herbig~Ae-like emission spectrum, infrared excess, and UXor-like light curve need another explanation, or it is much closer to us and to the Galactic plane. Dust-enshrouded stars in the Galactic halo are the R~Coronae Borealis stars, a few of them exhibiting early type spectra. These stars, however, are hydrogen-deficient, thus the spectrum of V1117~Her excludes this possibility. The disk around an evolved star might have originated from a disrupted close companion like in the case of BP~Psc (Zuckerman et~al. 2008).

\begin{table}[b]
{\small
\caption{Major parameters of the SED model plotted in Fig.~4.}
\begin{center}
\begin{tabular}{ccccccc}
\noalign{\smallskip}
\hline
\noalign{\smallskip}
$M_{*}$  & $T_{\mathrm eff}$ & Age  & $M_{\mathrm disk}$ &  $R^{\mathrm min}_{d}$  & $R^{\mathrm max}_{d}$ &  Incl. \\
(M$_{\odot}$) & (K)  & (Myr) & (M$_{\odot}$) & (AU) & (AU) & ($^{\mathrm o}$) \\
\noalign{\smallskip}
\hline
\noalign{\smallskip}
 1.95 &  7540  &  7.3  &  6.86E$-3$ &  0.3 &  8000  & 87  \\
\hline
\noalign{\smallskip}
\end{tabular}
\end{center}}
\end{table}

V1117~Her may be closer to the star-forming disk of the Milky Way if its circumstellar disk is viewed edge-on. The UXor phenomenon requires high inclination. It was demonstrated by De~Marchi et~al. (2013), that a disk inclination larger than $\sim 85^{\mathrm o}$ would reduce the brightness at optical wavelengths by some 4.5~mag. In this case the star's photosphere is completely occulted by its outer disk, and the whole optical flux is due to scattered light from the outer disk atmosphere, similar in colour to the unreddened star. In this case the observed brightness would indicate a distance of $\sim 300$\,pc and $z \approx 150$\,pc for V1117~Her. To explore this possibility further, we examined the SED model grid of Robitaille et~al. (2007). We looked for models that fulfil the following constraints, set by the observations. (1) The temperature of the central star is between 7300 and 7600\,K; (2) the age of the central star exceeds $10^6$ years (an assumption set by the absence of a star-forming environment); (3) the inner edge of the disk is located at the dust sublimation radius. We found that only one model, No.~3013085, viewed at an inclination of $87^{\mathrm o}$ and from a distance of 300\,pc produced an SED close to the observed SED of V1117~Her over the wavelength interval covered by observational data. The model is plotted in Fig.~4 with dash-dotted line. At longer wavelengths the model predicts increasing fluxes, however lower than the threshold of the {\it IRAS} at 60\,$\mu$m. The major parameters of the model are shown in Table~2. We conclude that a Herbig~Ae star similar in brightness and colour indices to V1117~Her can be located as close as 300\,pc to us. The age of the star, suggested by the modelling, is still higher than the typical lifetime of accretion disks, and at the hypothetical distance of 300\,pc the star is farther away from the Galactic plane than typical star forming regions.

\IBVSfig{7cm}{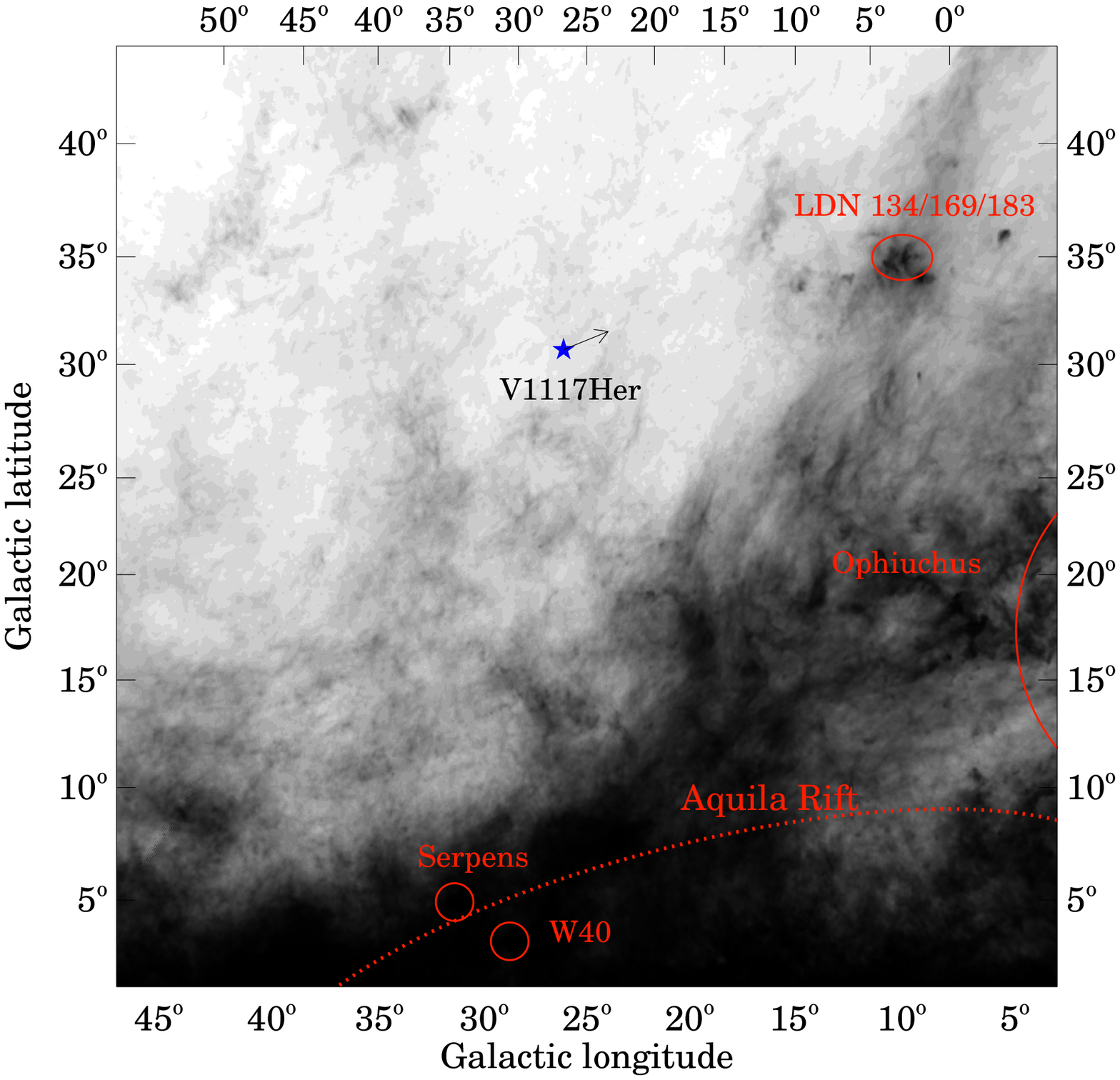}{Galactic environment and proper motion 
of V1117~Her. The distribution of the interstellar extinction in a field of 
$50\times50$~deg, centred on $l=25^{\mathrm o}$, $b=+25^{\mathrm o}$ is shown (Schlegel et~al. 1998). The star forming regions found in the field are indicated.}
\IBVSfigKey{6089-f5.ps}{V1117}{other}

Assuming a mass about 2\,M$_{\odot}$ for V1117~Her, the 408~day period in its brightness variations is probably caused by obscuring dust clumps orbiting the star at $\sim 1.35$\,AU. The amplitude and duration of the dimming, estimated from the shape of folded light curve in Fig.~2, allow us to estimate the mass of the clump. To this end we assume that V1117~Her has a primordial disk, with interstellar dust to gas ratio. The amplitude of 2.4\,mag then corresponds to 
$\Sigma \sim 0.007$\,g\,cm$^{-2}$ gas+dust column density (G\"uver \& \"Ozel 2009). The duration of the dip is about 25~days, which implies a clump size of 0.5\,AU at 1.35\,AU from the star. The mass of such a clump is some $6\times10^{-9}$\,M$_{\odot}$. Since the period persisted during several years, the clump must be dynamically fairly stable, possibly a proto-planetesimal, similar to the structure found by Chen et~al. (2012) in the circumstellar environment of GM~Cep. Fig.~1 suggests that the dips are deeper in the latest years than earlier, suggesting the growth of the clump.

Figure~5 shows the distribution of the interstellar dust (Schlegel et~al. 1998) in the wide-field Galactic environment of V1117~Her. The arrow indicates the displacement of the star during 1 million years, according to the proper motion given in the {\it NOMAD\/} Catalogue. At lower latitudes the Serpens, W40, and Aquila Rift star forming regions are located at 250--400\,pc. V1117~Her might have escaped from a lower latitude nearby star forming region. A mean velocity of 20\,km\,s$^{-1}$ is sufficient for travelling 150\,pc during 7~million years. At a distance of 300\,pc, the catalogued proper motions correspond to a tangential velocity of $\sim$14\,km\,s$^{-1}$, compatible with this scenario.
Reliable measurements of parallax, proper motion, and radial velocity, expected from Gaia, will certainly solve the contradictions in the observed properties of this star. 

\bigskip

{\bf Acknowledgements}
This work was supported by the Hungarian OTKA grant K81966 and the ESTEC Contract No.\,4000106398/12/NL/KML. The photometric contribution of {\it AAVSO} observers Teofilo Arranz, Laurent Bichon, Thomas Grzybowski, James McMath, Kenneth Menzies, Gary Poyner, Michael Simonsen is gratefully acknowledged. This research has made use of the VizieR catalogue access tool, CDS, Strasbourg, France. The authors are grateful to the referee for the thorough and  very helpful report.

\bigskip

\references

Ahn, C. P. et~al. 2012, {\it ApJS}, {\bf 203}, 21  

Akerlof, C. et~al. 2000, {\it  AJ}, {\bf 119}, 1901  

Bessell, M. S., \& Brett, J. M. 1988, {\it PASP}, {\bf 100}, 1134

Bibo, E. A., \&  Th\'e, P. S. 1990, {\it A\&A}, {\bf 236}, 155 \BIBCODE{1990A&A...236..155B}

Blazhko, S. 1929, {\it  Astron. Nachr.}, {\bf 236}, \nobibcode 279 \BIBCODE{1929AN....236..279W}

Chen, W. P. et~al. 2012, {\it ApJ}, {\bf 751}, 118

Cutri, R. M., et~al. 2003, VizieR On-line Data Catalog: II/246

De Marchi, G., Panagia, N., Guarcello, M. G., \& Bonito, R. 2013, {\it  MNRAS}, {\bf 435}, 3058

Dullemond, C. P. \& Monnier, J. D. 2010, {\it ARA\&A}, {\bf 48}, 205 

G\"uver, T., \& \"Ozel, F. 2009, {\it MNRAS}, {\bf 400}, 2050

Herbst, W., Herbst, D. K., Grossman, E. J., \& Weinstein, D. 1994, {\it AJ}, {\bf 108}, 1906 

Ishihara, D. et~al. 2010, {\it A\&A}, {\bf 514}, A1  

Ive\'zi\v{c}, \v{Z}. et~al. 2007, {\it ASPC}, {\bf 364}, \nobibcode 1651 \BIBCODE{2007ASPC..364..165I}

Jacoby, G. H., Hunter, D. H., \&  Christian, C. A. 1984, {\it ApJS\/}, {\bf 56}, 257

Kazarovets, E. V., Samus, N. N., Durlevich, O. V., Kireeva, N. N., Pastukhova, E. N. 2008, {\it IBVS}, {\bf 5863}, 1

Kenyon, S. J., \& Hartmann, L. 1995, {\it ApJS}, {\bf 101}, 117

Koll\'ath, Z. 1990, The Program Package MUFRAN, Occasional Technical Notes 
of the Konkoly Observatory, Budapest, No.\,1; \newline
 {\tt http://www.konkoly.hu/Mitteilungen/muf.tex}

Meyer, M. R., Calvet, N., \& Hillenbrand, L. A. 1997, {\it AJ}, {\bf 114}, 288

Natta, A., Prusti, T., Neri, R., Wooden, D., Grinin, V. P. \& Mannings, V. 2001,  {\it A\&A}, {\bf 371}, 186

Pecaut, M. J., \& Mamajek, E. E. 2013, {\it ApJS}, {\bf 208}, 9

Pojmanski, G.  2002, {\it Acta Astronomica}, {\bf 52}, 397 \BIBCODE{2002AcA....52..397P}

Robitaille, T. P., Whitney, B. A., Indebetouw, R., \& Wood, K. 2007, {\it ApJS}, {\bf 169}, 328

Rostopchina, A.N., Grinin, V.P., \& Shakhovskoi, D.N. 1999, {\it Astr. Lett.}, {\bf 25}, 243

Samus, N. et~al. 2007-2012, {\it General Catalogue of Variable Stars\/},  VizieR On-line Data Catalog: B/gcvs
	
Schlegel, D. J., Finkbeiner, D. P., \& Davis, M. 1998, {\it ApJ}, {\bf 500}, 525

Siess, L., Dufour, E., Forestini, M. 2000, {\it A\&A}, {\bf 358}, 593 

Wo\'zniak, P. R. et~al. 2004a, {\it AJ}, {\bf 127}, 2436  

Wo\'zniak, P. R., Williams, S. J., Vestrand, W. T., Gupta, V. 2004b, {\it AJ}, {\bf 128}, 2965  

Wright, E. L. et~al. 2010, {\it AJ}, {\bf 140}, 1868  

Xiao, L., Kroll, P., \& Henden, A. A. 2010, {\it AJ},  {\bf 139}, 1527 

Zuckerman, B. et~al. 2008, {\it ApJ}, {\bf 683}, 1085

\endreferences

\end{document}